\date{}
\documentclass[showpacs]{revtex4}
\usepackage{graphicx,psfrag,amsmath,amssymb,amsfonts,bbm,latexsym,color,dcolumn,epsf}
\begin{document}
\title{Quantum dissipative effects in moving mirrors: a functional approach}
\author{C. D. Fosco$^a$ }
\author{F. C. Lombardo$^b$ }
\author{F. D. Mazzitelli$^b$ }
\affiliation{$^a$Centro At\'omico Bariloche and Instituto Balseiro,
 Comisi\'on Nacional de Energ\'\i a At\'omica, \\
R8402AGP Bariloche, Argentina}
\affiliation{$^b$Departamento de
F\'\i sica {\it Juan Jos\'e Giambiagi}, FCEyN UBA, Facultad de
Ciencias Exactas y Naturales, Ciudad Universitaria, Pabell\' on I,
1428 Buenos Aires, Argentina.}
\date{today}
%====================================================================
\begin{abstract}
\noindent We use a functional approach to study various aspects of
the quantum effective dynamics of moving, planar, dispersive
mirrors, coupled to scalar or Dirac fields, in different numbers
of dimensions. We first compute the Euclidean effective action,
and use it to derive the imaginary part of the `in-out' effective
action.
We also obtain, for the case of the real scalar field in $1+1$ dimensions,
the Schwinger-Keldysh effective action and a semiclassical Langevin
equation that describes the motion of the mirror including
noise and dissipative effects due to its coupling to the quantum
fields.
\end{abstract}
%%%%%%%%%%%%%%%%%%%%%%%%%%%%%%%%%%%%%%%%%%%%%%%%%%%%%%%%%%%%%%%%%%%%%
%%%%%%%%%%%%%%%%%%%%%%%%%%%%%%%%%%%%%%%%%%%%%%%%%%%%%%%%%%%%%%%%%%%%%
%%%%%%%%%%%%%%%%%%%%%%%% Introduction %%%%%%%%%%%%%%%%%%%%%%%%%%%%%%%
%%%%%%%%%%%%%%%%%%%%%%%%%%%%%%%%%%%%%%%%%%%%%%%%%%%%%%%%%%%%%%%%%%%%%
%%%%%%%%%%%%%%%%%%%%%%%%%%%%%%%%%%%%%%%%%%%%%%%%%%%%%%%%%%%%%%%%%%%%%
\maketitle
\section{Introduction}\label{sec:intro}
In the presence of a moving, accelerated mirror, the
electromagnetic field evolves from the vacuum to an excited state,
containing a non-vanishing number of photons. This `motion induced
radiation' or `Dynamical Casimir Effect' (DCE) has been the
subject of intense theoretical research since its discovery in the
seventies~\cite{moore,fv}. While this phenomenon was initially
regarded as being of just theoretical interest (for example as a
toy model for black hole evaporation), in recent years it has been
pointed out that the experimental verification of the DCE might
not be, after all, so far out of reach~\cite{job,Dodonov-rev}.

Indeed, taking advantage of parametric resonance amplification,
this effect could be dramatically increased \cite{varios}, since
the number of photons created within a cavity with a moving mirror
should grow exponentially at resonance (i.e., when the mirror's
oscillatory frequency doubles one of the eigenfrequencies of the
unperturbed cavity). For the case of microwave cavities, the
mechanical frequency of the mirror should, however, be extremely
high ($\sim$ 1GHz) for this to happen, and this poses the main
stumbling block for an experimental verification of the effect.

It has also been suggested that the DCE could be measured in
experiments in which the moving mirror is replaced by a
semiconductor slab which suddenly changes its conductivity due to
illumination with short laser pulses~\cite{braggio}.
Unfortunately, the unavoidable losses in the semiconductor could
put the viability of this proposal in jeopardy~\cite{dodloss}. Yet
another alternative that has been advanced~\cite{onofrioprl},
which amounts to consider an array of nanoresonators, moving
coherently at frequencies in the GHz range. The detection of the
created photons could, in this case, be performed using an
inverted population of Rydberg atoms.

From the theoretical point of view, the DCE has been analyzed for a variety
of geometries and using many different theoretical tools. A particularly
interesting functional approach has been proposed by Golestanian and
Kardar~\cite{GK}. They introduce auxiliary fields in the
functional integral for the quantum field, whose role is to impose
the boundary conditions on the mirrors. This method has been successfully
applied, for example, to the calculation of the Euclidean effective action
for one and two (slightly deformed) moving mirrors in $d+1$
dimensions~\cite{miri}, deriving also the effective equation of motion for
the mirror by analytic continuation of the Euclidean effective action.

In view of the possibility of detecting the DCE using
nanoresonators~\cite{onofrioprl}, it is of interest to extend this
formalism in several directions. On the one hand, it is important
to generalize the method, in order to be able to consider
dispersive mirrors, rather than just perfectly conducting ones. On
the other hand, since the nanoresonators could eventually show
quantum behaviour~\cite{quantumnano}, it is worthwhile to consider
their quantum to classical transition, and to describe their
effective dynamics in terms of a semiclassical Langevin equation.
This paper is a step in that direction~\cite{previousctp}.

Besides, to exhibit the quite general nature of the phenomenon, it is also
interesting to extend the formalism to consider mirrors coupled to
different fields, like the case of a moving wall that imposes bag
conditions on a Dirac field. In this article, we first show how to
generalize the functional approach of~\cite{GK}, to calculate the Euclidean
effective action for moving dispersive mirrors coupled to real scalar and
then to Dirac fields. We also show how the case of a relativistic mirror
also fits in the formalism, by performing minor modifications.
We then compute the Schwinger-Keldysh or Closed Time Path (CTP) effective
action for a mirror coupled to a scalar field. More realistic situations
(a cavity with two no-flat mirrors coupled to the electromagnetic field)
will be considered in a forthcoming publication.

This article is organized as follows: in section~\ref{sec:effect},
we use a path-integral approach to evaluate the Euclidean
effective action for a single, perfect or imperfect,
non-relativistic moving mirror in $1+1$ dimensions, both for the
real scalar and Dirac field cases. By `perfect mirror' we mean one
that imposes Dirichlet boundary conditions, when coupled to a
scalar field, or bag conditions in the Dirac field case. In both
cases, the boundary conditions due to the perfect mirror are
introduced by the coupling of the quantum field to a singular mass
term, localized on a region of codimension $1$, with a divergent
coupling constant $\lambda \to \infty$.  The imperfect mirrors
that we shall consider here will be, on the other hand, described
by the same kind of interaction term, albeit with a {\em finite\/}
coupling constant $\lambda$.

The changes needed to cope with the relativistic mirror generalization are
also presented, taking the real scalar field case as a concrete example,
and evaluating the corresponding effective action.

In section~\ref{sec:dpo}, we consider the case of a (flat)
moving mirror in $d+1$ dimensions, coupled to a real scalar field,
evaluating explicitly the Euclidean effective action. In
section~\ref{sec:imaginary}, we evaluate and interpret the
imaginary part of the in-out effective action, obtained after Wick
rotating to real time, for the case of the real scalar field with
perfect boundary conditions.

In section~\ref{sec:eqs}, we evaluate the quantum corrections to the
mirrors' real-time equations of motion. In order to do this, we
compute the CTP effective action and obtain a semiclassical
stochastic equation for the mirror. Moreover, from the imaginary
part of the CTP effective action we provide an estimation of the
decoherence time for the mirror.

\section{Moving mirrors in $1+1$ dimensions}\label{sec:effect}
\subsection{Real scalar field}\label{ssec:scalar2}
We shall begin by considering a massive real scalar field $\varphi$
coupled to an imperfect mirror, whose position is described by a
function $q(x^0)$, so that the real-time Lagrangian density,
${\mathcal L}$ is:
\begin{equation}\label{eq:defls}
{\mathcal L} \;=\; \frac{1}{2} \partial_\mu\varphi \partial^\mu \varphi
-\frac{1}{2} m^2 \varphi^2 - \frac{1}{2} \, V(x^0,x^1) \,\varphi^2 \;,
\end{equation}
where $V$ is a $\delta$-like singular function:
\begin{equation}\label{eq:defsing}
V(x^0,x^1) \,=\, \lambda \, \delta(x^1 - q(x^0))\;,
\end{equation}
determined by the mirror's position. $\lambda$ is a positive
coupling constant. The coupling to the singular field has the
effect of introducing a perfect mirror (at $x^1 = q(x^0)$) when
$\lambda \to + \infty$, since it then enforces the condition
$\varphi = 0$ on the points of the spacetime curve ${\mathcal C}$
defined by the points $(x^0,\, q(x^0))$.  On the other hand, the
imperfect mirror situation is simulated for $0 < \lambda < \infty$
\cite{barton1}.

Let us now perform a Wick rotation: $x^0 = - i \tau$, and
calculate the resulting Euclidean  effective action
$\Gamma[q(\tau)]$ for the mirror, due to the scalar-field vacuum
fluctuations, in the functional-integral representation:
\begin{equation}\label{eq:defzsc}
e^{ - \Gamma[q(\tau)]} \,\equiv\, {\mathcal Z}[q(\tau)] \,=\,
    \int {\mathcal D} \varphi \; e^{ - S[\varphi;q] } \;,
    \end{equation}
    where $S[\varphi;q]$ is the Euclidean action:
    \begin{equation}
    S[\varphi;q] \;=\; S_0[\varphi] \;+\; S_{\mathcal C}[\varphi; q]
    \end{equation}
    with $S_0$ denoting the free part
    \begin{equation}
    S_0[\varphi]\,=\, \frac{1}{2} \, \int d^2x \big( \partial_\mu
\varphi\partial_\mu
            \varphi  + m^2 \varphi^2 \big)
    \end{equation}
and $S_{\mathcal C}$ the coupling to the mirror,
\begin{equation}
S_{\mathcal C}[\varphi; q] \,=\, \frac{\lambda}{2} \, \int d^2x
 ~\delta(x_1 - q(x_0)) \big[ \varphi(x) \big]^2 \,=\,
\frac{\lambda}{2}\, \int d\tau \big[\varphi(\tau,q(\tau))]^2 \;.
\end{equation}
Euclidean coordinates are denoted by $x_\mu$, where $x_0 \equiv
\tau$; the metric tensor is the $2\times 2$ identity matrix.

To proceed, we introduce an auxiliary field $\xi(\tau)$, living in $0+1$
dimensions, whose role is to linearize the term $S_{\mathcal C}$, which
couples the scalar field to the mirror. The resulting expression for
${\mathcal Z}[q(\tau)]$ is:
\begin{equation}
{\mathcal Z}[q(\tau)] \;=\;\int {\mathcal D}\xi \; e^{-\frac{1}{2
\lambda} \int d\tau \xi^2(\tau)} \; {\mathcal Z}_0 [J_\xi]
\label{auxfield}
\end{equation}
where ${\mathcal Z}_0$ is the free generating functional:
\begin{equation}
 {\mathcal Z}_0[J] \,=\, e^{- W_0 [J]} \,=\, \int {\mathcal D} \varphi \;
e^{- S_0(\varphi)  + i \int d^2 x J(x) \varphi(x)} \;,
\end{equation}
and $J_\xi$ is a current localized on the defect and proportional
to the auxiliary field: \mbox{$J_\xi(x_0,x_1) \equiv \xi(x_0)
\,\delta(x_1 - q(x_0))$}. Note that (\ref{auxfield}) reduces to
the approach of Golestanian and Kardar \cite{GK} when $\lambda \to
\infty$, i.e., when a perfect mirror is considered.

Since the integral over $\varphi$ is a Gaussian, we can
immediately write down the explicit form of $W_0$,
\begin{equation}
W_0[J] \;=\; \frac{1}{2} \int d^2x \int d^2 x' J(x) \Delta (x-x') J(x') \;,
\end{equation}
where $\Delta$ is the free Euclidean correlation function:
\begin{eqnarray}
\Delta(x-y) &=& \langle \varphi(x) \varphi(y)\rangle
\nonumber\\
&=& \int \frac{d^2k}{(2\pi)^2} e^{-i k \cdot (x-y)} \frac{1}{ k^2 + m^2} \;.
\end{eqnarray}
Thus we derive for ${\mathcal Z}[q(\tau)]$ a `dimensionally
reduced' path integral expression involving  just the auxiliary
field $\xi$,
\begin{equation}
{\mathcal Z}[q(\tau)] \,=\, \int {\mathcal D}\xi \; e^{-\frac{1}{2} \int
d\tau \int d\tau' \xi(\tau) {\mathcal K}(\tau,\tau') \xi(\tau')} \;,
\end{equation}
where we have introduced the kernel $K(\tau,\tau')$
\begin{equation}
{\mathcal K}(\tau,\tau') \;=\; \frac{1}{\lambda} \delta(\tau-\tau') +
\Delta\big[\tau-\tau',q(\tau)-q(\tau')\big] \;.
\end{equation}
The $\xi$-integral, again a Gaussian, allows as to write down the
(formal) result for ${\mathcal Z}[q(\tau)]$ as follows:
\begin{equation}
{\mathcal Z}[q(\tau)] \;=\; \big( \det {\mathcal K} \big)^{-\frac{1}{2}}
\end{equation}
so that
\begin{equation}\label{eq:gqdet}
\Gamma[q(\tau)] \;=\; \frac{1}{2} {\rm Tr} \big[\ln {\mathcal K} \big] \;.
\end{equation}
Let us now approximate (\ref{eq:gqdet}) for small departures with
respect to the static mirror case. To that end, we first expand
${\mathcal K}$:
\begin{equation}
{\mathcal K}  \;=\; {\mathcal K}_0 \,+\, {\mathcal K}_1 \,+\, {\mathcal
K}_2 \,+\, \ldots
\end{equation}
where the subscripts denote the order of the corresponding term.
To derive the linearized form of the equations of motion, it shall
be sufficient to keep terms of up to the quadratic order. It is
quite straightforward to see that
\begin{equation}
{\mathcal K}_0(\tau,\tau')\;=\;\int \frac{d\omega}{2\pi} e^{i \omega
(\tau-\tau')} {\widetilde K}_0 (\omega) \;,\;\;\;
{\widetilde{\mathcal K}}_0 (\omega) \;=\; \frac{1}{\lambda}
\,+\, \frac{1}{2\sqrt{\omega^2 + m^2}} \;,
\end{equation}
\begin{equation}
{\mathcal K}_2 (\tau ,\tau') \;=\; \frac{1}{4} \big(q(\tau) -
q(\tau')\big)^2 \, \int \frac{d\omega}{2\pi} \, e^{i \omega(\tau-\tau')}
\sqrt{\omega^2 + m^2} \;,
\end{equation}
and that ${\mathcal K}_1$ vanishes. It should be kept in mind that $q(\tau)$ is
the {\em departure\/} with respect to a constant (fixed to $0$ by a shift
of the axis, if necessary). This implies, in particular, that its Fourier
transform ${\tilde q}(\omega)$ will verify ${\tilde q}(0)=0$. Of course,
${\tilde q}(0)=0$ alone does not imply a small departure. Indeed, the
condition holds true for some motions that correspond to an unbounded
motion, like ${\tilde q}(\omega) \propto i \delta^{'''}(\omega)$, which
comes from $q(\tau) \propto \tau^3$. But in this case the quadratic
approximation fails, since $q(\tau)$ becomes large (and ${\tilde
q}(\omega)$ singular).

Coming back to the expression for $\Gamma[q(\tau)]$, expanding up to the
second order in the fluctuation, and discarding a $q(\tau)$-independent
term, we see that
\begin{equation}
\Gamma[q(\tau)] \,=\, \frac{1}{2} \,{\rm Tr} \ln \big[{\mathcal K}_0 + {\mathcal K}_2 \big]
\, \simeq \,  \frac{1}{2} {\rm Tr} \big[{\mathcal K}_0^{-1} {\mathcal K}_2 \big]
\,\equiv\, \Gamma_2[q(\tau)]\;,
\end{equation}
where $\Gamma_2[q(\tau)]$ may be written more explicitly as follows:
\begin{equation}
\Gamma_2[q(\tau)] \,=\, \frac{1}{2} \int_{-\infty}^{+\infty} d\tau \int_{-\infty}^{+\infty}  d\tau'
\big[{\mathcal K}_0^{-1}(\tau,\tau') {\mathcal K}_2(\tau',\tau)\big] \;.
\end{equation}

Using the explicit form for ${\mathcal K}_0$ and ${\mathcal K}_2$,
\begin{equation}
\Gamma_2[q(\tau)] \,=\, \frac{1}{2} \int_{-\infty}^{+\infty} d\tau \int_{-\infty}^{+\infty}
d\tau' ( q(\tau) - q(\tau') )^2  F(\tau -\tau') \;,
\end{equation}
where
\begin{equation}
  F(\tau -\tau') \,=\, \int \frac{d\omega}{2\pi} e^{i \omega(\tau-\tau')}
{\tilde F}(\omega) \,,
\end{equation}
with
\begin{equation}\label{eq:defft}
{\tilde F}(\omega) \,=\,\frac{1}{4} \, \int \frac{d\nu}{2\pi} \,
\big[\frac{1}{\lambda} \,+\, \frac{1}{2 \sqrt{(\nu + \omega)^2 + m^2}}\big]^{-1}
\sqrt{\nu^2 + m^2}\;.
\end{equation}
It is clear that we may subtract from ${\tilde F(\omega)}$ its
value at zero-frequency, since any $\omega$-independent part would
give zero when inserted in $\Gamma_2[q(\tau)]$ (it would produce a
$\delta(\tau-\tau')$ contribution to $F$, multiplied by a
continuous function that vanishes when $\tau = \tau'$). Thus we
introduce
\begin{equation}
{\tilde F}_s(\omega) \,\equiv\, {\tilde F}(\omega) \,-\, {\tilde F}(0) \,,
\end{equation}
the subtracted version of ${\tilde F}$. Since ${\tilde F}_s(0) = 0$, we
obviously have \mbox{$\int d\tau F_s(\tau) = 0$}, and
\begin{eqnarray}
\Gamma_2[q(\tau)] &=& \frac{1}{2} \int_{-\infty}^{+\infty} d\tau \int_{-\infty}^{+\infty}  d\tau'
( q(\tau) - q(\tau') )^2  F(\tau -\tau') \nonumber\\
&=& \frac{1}{2} \int_{-\infty}^{+\infty} d\tau \int_{-\infty}^{+\infty}  d\tau'
( q(\tau) - q(\tau') )^2  F_s(\tau -\tau') \nonumber\\
 &=& - \int_{-\infty}^{+\infty} d\tau \int_{-\infty}^{+\infty}  d\tau'
q(\tau) q(\tau') \, F_s(\tau -\tau') \;. \label{eea}
\end{eqnarray}

Expression (\ref{eq:defft}) for ${\tilde F}$ is divergent; to
regulate it we introduce a symmetric frequency cutoff $\Xi$, such
that $|\nu | \leq \Xi$, and the regulated function ${\tilde
F}_s(\omega,\Xi)$ is:
\begin{eqnarray}\label{eq:deffto}
{\tilde F}_s(\omega,\Xi) &=&
\frac{1}{4} \, \int_{-\Xi}^\Xi \frac{d\nu}{2\pi} \,\Big\{
\big[\frac{1}{\lambda} \,+\, \frac{1}{2 \sqrt{(\nu + \omega)^2 + m^2}}\big]^{-1}
\sqrt{\nu^2 + m^2} \nonumber\\
&-& \big[\frac{1}{\lambda} \,+\, \frac{1}{2 \sqrt{\nu^2 + m^2}}\big]^{-1}
\sqrt{\nu^2 + m^2} \Big\} \;.
\end{eqnarray}
${\tilde F}_s(\omega,\Xi)$ in (\ref{eq:deffto}) is convergent for
$\Xi \to \infty$, so this `symmetric-limit' regularization  yields
a finite value for \mbox{${\tilde F}_s(\omega)\equiv \lim_{\Xi \to
\infty}{\tilde F}_s(\omega,\Xi)$} when the regulator is removed.
Unfortunately, there seems to be no analytic expression for
${\tilde F}_s(\omega)$ which is valid for arbitrary values of the
constants $m$ and $\lambda$. We can, however, calculate it for
different relevant particular cases:
\begin{enumerate}
\item $m = 0$, $\lambda \to \infty$: in this case, we have
\begin{eqnarray}\label{eq:f0i}
{\tilde F}_s(\omega) &=& \lim_{\Xi \to \infty}
\frac{1}{2} \, \int_{-\Xi}^\Xi \frac{d\nu}{2\pi} \,
\Big[ |\nu + \omega| |\nu| \,-\,|\nu|^2 \Big] \nonumber\\
 &=& \frac{1}{12\pi}  |\omega|^3 \;.
\label{fsperf}
\end{eqnarray}
A cubic dependence in $\omega$ could have been guessed on
dimensional grounds. The numerical coefficient coincides with
previously obtained results \cite{LR}.
\item $m = 0$, $\lambda < \infty$: a property that we immediately see is that,
for any finite $\lambda$, the large-$\nu$ behaviour of the integral is
improved (by a power of $\nu$) with respect to the perfect mirror
($\lambda \to \infty$) case. As a consequence, the result obtained by taking the
$\Xi \to \infty$ and $\lambda \to \infty$ limits will depend on the order in
which they are taken.

The $\nu$ integral (for a finite $\lambda$) and its $\Xi \to
\infty$ limit can be evaluated exactly in this case, the result
being,
\begin{equation}
{\tilde F}_s(\omega) \,=\, \frac{\lambda^2}{16\pi^2} \Big[
2 |\omega| - \lambda \big( 1 + \frac{2}{\lambda} |\omega| \big)
\ln\big(1 + \frac{2}{\lambda} |\omega|\big)\Big]\;.
\end{equation}
Performing a large-$\lambda$ expansion in the previous expression;
we see that
\begin{equation}\label{eq:flexp}
{\tilde F}_s(\omega) \,=\, - \frac{\lambda}{8\pi} \omega^2
\,+\, \frac{1}{12\pi}  |\omega|^3 \,+\,{\mathcal O}(\lambda^{-1})\;.
\end{equation}
The second term is independent of $\lambda$, and it coincides with
the result Eq.(\ref{fsperf}) obtained for $\lambda \to \infty$.
The first term was absent from the perfect mirror case, and is a
reflection of the fact that, as anticipated, the $\Xi \to \infty$
and $\lambda \to \infty$ limits do not commute. The reason for
that difference is that the finite-$\lambda$ system includes the
effect of more quantum fluctuations than in the infinite-$\lambda$
case. The resulting difference between the results obtained for
those different limits has, however, a simple physical
interpretation. Indeed, that difference $\delta{\tilde F}_s$ comes
from the ${\mathcal O}(\lambda)$ term in Eq.(\ref{eq:flexp}):
\begin{equation}
\delta{\tilde F}_s(\omega) \,\equiv \, - \frac{\lambda}{8\pi} \omega^2\,,
\end{equation}
a term which, when inserted into the expression for $\Gamma_2[q]$ yields
\begin{equation}
\delta\Gamma_2[q] \,=\, \int d\tau \frac{1}{2} \, \mu(\lambda)
\dot{q}^2(\tau) \;,
\end{equation}
where $\mu(\lambda) = \frac{\lambda}{4\pi}$. This term can, of
course, be regarded as a renormalization in the mirror's mass,
when the mirror has a non-relativistic kinetic-energy term, as it
is usually assumed. It shouldn't come as a surprise that the
outcome of the calculation is a non-relativistic invariant object:
the coupling between mirror and field, $S_{\mathcal C}$, does in
fact assume a non-relativistic description for the mirror, since
it is not a relativistic invariant. The covariant formulation of
this example is presented, for the sake of completeness, in
~\ref{ssec:cova}.

\item $m \neq 0$, $\lambda \to \infty$: the exact result for this
case can also be obtained, although the calculation is more
involved. As outlined in Appendix A, the final result is:
\begin{equation}
\tilde F_s(\omega)=\frac{1}{12\pi}\int_0^1\frac
{d\alpha}{\alpha(\alpha-1)}\left [\left [\alpha(\alpha
-1)\omega^2+m^2\right]^{3/2}-m^3\right]
\label{Fmperf}
\end{equation}
which reduces to the proper result Eq.(\ref{fsperf}) in the $m \to
0$ limit.
\end{enumerate}

\subsection{Real scalar field: relativistic mirror}\label{ssec:cova}
 We present here a relativistically invariant formulation of the real
scalar field case.
 The main reason, besides its intrinsic interest, is that
 it makes it easier to understand the approximation incurred in the non relativistic
approach we (implicitly) used in the previous subsections. This
problem has been considered previously in Ref.~\cite{barton2} using
a canonical formalism.

An explicitly invariant coupling can be constructed, for example,
by considering a relativistic generalization of coupling term,
$S_{\mathcal C} \to S^{\rm rel}_{\mathcal C}$:
\begin{equation}
S^{\rm rel}_{\mathcal C}[\varphi,q] \;=\; \frac{\lambda}{2} \,
\int d^2x \int ds  \sqrt{\dot{q}_\mu(s)\dot{q}_\mu(s)} \,
\delta^{(2)}[x - q(s)] \big[ \varphi(x) \big]^2
\end{equation}
where $q_\mu(s)$, $\mu=0,1$,  is a suitable parametrization of the
worldline described by the mirror. We use the notation $\dot{q}_\mu=
\frac{dq_\mu}{ds}$.

When the parametrization is such that
$s$ coincides with the laboratory time, $x_0 \to (x_0, q(x_0))$, we obtain:
\begin{equation}
S^{\rm rel}_{\mathcal C}[\varphi,q] \;=\; \frac{\lambda}{2} \,
\int dx_0 \sqrt{1 + \dot{q}^2(x_0)} \,
\big[ \varphi(x_0, q(x_0)) \big]^2 \;,
\end{equation}
which indeed reduces to the non-relativistic term $S_{\mathcal C}$ for
\mbox{$|\dot{q}| << 1$}, and justifies {\em a posteriori\/} the non-relativistic
coupling when that condition is fulfilled.

Let us now write down the expressions for the (Euclidean) relativistic versions
of the objects we have considered before:
\begin{equation}
{\mathcal Z}^{\rm rel}[q(s)] \;=\;\int {\mathcal D}\xi \; e^{-\frac{1}{2 \lambda} \int
ds \xi^2(s)} \; {\mathcal Z}_0[J_\xi^{\rm rel}]
\end{equation}
where
\begin{equation}
J_\xi^{\rm rel} (x_0,x_1) \equiv \int ds \xi(s) \, |\dot{q}(s)|^{\frac{1}{2}}
\delta^{(2)}(x - q(s)) \;.
\end{equation}
Then
\begin{equation}
e^{-\Gamma^{\rm rel}[q(s)]} = {\mathcal Z}^{\rm rel}[q(s)] \;=\;
\big( \det {\mathcal K}^{\rm rel}\big)^{-\frac{1}{2}}
\end{equation}
where
\begin{equation}
{\mathcal K}^{\rm rel}(s,s')\;=\;\frac{1}{\lambda} \,\delta(s-s') \,+\,
|\dot{q}(s)|^{\frac{1}{2}} \,\Delta\big[q(s)-q(s')\big] \,|\dot{q}(s')|^{\frac{1}{2}} \,.
\end{equation}

The next step is to perform an expansion in powers of $f_\mu(s)$, the fluctuating
part of $q_\mu(s)$:
\begin{equation}
q_\mu(s) \;=\; q_\mu^{(0)}(s) \,+\, f_\mu(s) \;
\end{equation}
where $q_\mu(s)$ is analogous to the worldline for the `static' mirror.
It is in fact a linear function of $s$
\begin{equation}
q_\mu^{(0)} (s) \;=\; a_\mu \,+\, v_\mu \, s
\end{equation}
where $a_\mu$ and $v_\mu$ are constant vectors;  $v_\mu$ is
time-like in the real time description. The fluctuating part,
$f_\mu$, will be assumed to be such that $\dot{f}_\mu$ is
orthogonal to $v_\mu$. The reason is that parallel components
amount to fluctuations in the parametrization, which are of course
irrelevant in a reparametrization-invariant theory.

In order to simplify matters, we use in what follows a specific convariant
parametrization; namely, we assume that $s$ is the mirror's proper time.
Then the $|\dot{q}|^{\frac{1}{2}}$ factors become both equal to $1$.
The calculations then proceed, for this parametrization, in a way that
mimics the non-relativistic ones. The quadratic part of the effective
action is given by,
\begin{equation}
\Gamma_2^{\rm rel}[q(s)] \,=\, \frac{1}{2} \int_{-\infty}^{+\infty} ds \int_{-\infty}^{+\infty}  ds'
\big[({\mathcal K}^{\rm rel})_0^{-1}(s,s') {\mathcal K}^{\rm rel}_2(s',s)\big] \;.
\end{equation}
where
\begin{eqnarray}
{\mathcal K}^{\rm rel}_0(s,s')&=& \frac{1}{\lambda} \, \delta(s-s') \,+\,
\int \frac{d^2 k}{(2\pi)^2} e^{i k \cdot v (s-s')} \, \frac{1}{k^2 + m^2}
\nonumber\\
&=& \int \frac{d\omega}{2\pi}\, e^{i \omega (s-s')} \frac{1}{\lambda} \,+\,
\int \frac{d\omega}{2\pi}\, e^{i \omega |v|(s-s')}  \frac{1}{2 \sqrt{\omega^2 + m^2}}
\nonumber\\
&=& \int \frac{d\omega}{2\pi}\, e^{i \omega (s-s')} \Big[ \frac{1}{\lambda} \,+\,
\frac{1}{2 |v| \, \sqrt{\omega^2 + m^2}} \Big]
\end{eqnarray}
and
\begin{equation}
{\mathcal K}^{\rm rel}_2 (s,s') \;=\; \frac{1}{4} \big[f(s) - f(s')\big]^2 \,
\frac{1}{|v|} \, \int \frac{d\omega}{2\pi} \, e^{i \omega (s-s')} \,\sqrt{\omega^2 + m^2} \;.
\end{equation}
Then
\begin{equation}
\Gamma_2^{\rm rel}[q(s)] \;=\; - \,\int_{-\infty}^{+\infty} ds \int_{-\infty}^{+\infty}  ds'
q(s)  \, F_s^{\rm rel} (s -s') \, q(s') \;.
\end{equation}
where:
\begin{eqnarray}
{\tilde F}^{\rm rel}_s(\omega) &=& \lim_{\Xi \to \infty}
\frac{1}{4} \, \int_{-\Xi}^\Xi \frac{d\nu}{2\pi} \,\Big\{
\big[\frac{|v|}{\lambda} \,+\, \frac{1}{2 \sqrt{(\nu + \omega)^2 + m^2}}\big]^{-1}
\sqrt{\nu^2 + m^2} \nonumber\\
&-& \big[\frac{|v|}{\lambda} \,+\, \frac{1}{2 \sqrt{\nu^2 + m^2}}\big]^{-1}
\sqrt{\nu^2 + m^2} \Big\} \;.
\end{eqnarray}
This implies, in particular, that in the $\lambda \to \infty$ limit, the
results for $\Gamma^{\rm rel}[q]$ coincide with the ones for the non-relativistic
case, if one uses the proper-time as the evolution parameter.

It is interesting to consider now the situation when $\lambda$ is large but
finite and $m =0$. It is quite straightforward to see that this contribution
generates, as in the non-relativistic case,  an order-$\lambda$ term:
\begin{equation}
\delta\Gamma_2^{\rm rel}[q] \,=\, \frac{1}{|v|} \,\int ds \frac{1}{2} \, \mu(\lambda)
\dot{f}^2(s) \;,
\end{equation}
where $\mu(\lambda) = \frac{\lambda}{4\pi}$. On the other hand,
since \mbox{$\dot{q}^2(s)= v^2 + \dot{f}^2$} and we have an
equivalent expression for $\delta\Gamma_2^{\rm rel}$:
\begin{equation}
\delta\Gamma_2^{\rm rel}[q] \,=\, \mu^{\rm rel}(\lambda,v) \int ds \;,
\end{equation}
a term that has a form of a ($v$-dependent) mass counterterm for the relativistic mirror
action, with
\begin{equation}
\mu^{\rm rel}(\lambda,v) \;\equiv\;\frac{1}{2} \mu(\lambda) (\frac{1}{|v|}- |v|)
\;,
\end{equation}
since the natural relativistic action for a mirror with mass $M$ is:
\begin{equation}
S_{\rm mirror}^{\rm rel} \;=\; M \,\int ds \;.
\end{equation}

\subsection{Dirac field}\label{ssec:dirac}
Let us now consider the case of a Dirac field with bag-like
boundary conditions on the `mirror'. This kind of boundary
condition can also be introduced by means of an interaction with a
singular potential; indeed, as shown in~\cite{Sundberg:2003tc},
bag-like boundary conditions may be introduced by considering the
limit of a singular mass term. The real-time Lagrangian density is
then
\begin{equation}
{\mathcal L} \;=\; {\bar\psi} \big[ i \not \! \partial - m - V(x^0,x^1)
\big] \psi(x)
\end{equation}
where $V$ is the singular potential defined in (\ref{eq:defsing}).

As in the real scalar field case, we may pass to the Euclidean formulation,
to calculate $\Gamma[q(\tau)]$ for the mirror:
\begin{equation}
e^{-\Gamma[q(\tau)]}\;=\; {\mathcal Z}[q(\tau)] \;=\; \int {\mathcal D}\psi {\mathcal
D}{\bar\psi} \, e^{-S[{\bar\psi},\psi; q]} \;,
\end{equation}
where now:
\begin{equation}
S[{\bar\psi},\psi; q] \,=\, S_0[{\bar\psi},\psi] \,+\, S_{\mathcal C}[{\bar\psi},\psi; q]
\end{equation}
with
\begin{equation}
S_0[{\bar\psi},\psi] \,=\, \int d^2x ~{\bar\psi} ~ (\not \! \partial + m )
\psi
\end{equation}
and
\begin{equation}
S_{\mathcal C}[{\bar\psi},\psi;q] \,=\, \lambda \, \int d^2x ~{\bar\psi}(x)
~\delta[x_1 - q(x_0)] ~\psi(x) \;.
\end{equation}
Now to linearize the coupling we need two $2$-component (Grassmann) auxiliary fields,
${\xi}(\tau)$ and ${\bar\xi}(\tau)$, so that
\begin{equation}
{\mathcal Z}[q(\tau)] \;=\; \int {\mathcal D}\xi {\mathcal D}{\bar\xi} \,
e^{-\frac{1}{\lambda}\int d\tau {\bar\xi}(\tau) \xi(\tau)} \;
{\mathcal Z}_0[{\bar \eta}_\xi,\eta_\xi]
\;,
\end{equation}
with
\begin{equation}
{\mathcal Z}_0[{\bar\eta},\eta] \;=\;
e^{-W_0[{\bar\eta},\eta]}
\end{equation}
where
\begin{equation}
W_0[{\bar\eta},\eta] \;=\; \int d^2x \int d^2x' ~{\bar\eta}(x)
~{\mathcal S}_f (x-x') ~\eta(x')
\end{equation}
and ${\mathcal S}_f$ is the free Dirac propagator:
\begin{equation}
{\mathcal S}_f(x,x') \,=\, \langle \psi(x) {\bar\psi}(x') \rangle \,=\,
\int \frac{d^2p}{(2\pi)^2} e^{i p \cdot (x-x')} \frac{1}{i \not\! p + m} \;.
\end{equation}
We have introduced the sources:
\begin{equation}
\eta_\xi(x) \,=\, \xi (x_0) \, \delta[x_1 - q(x_0)] \;\;,\;\;\;
{\bar\eta}_{\bar\xi}(x) \,=\, {\bar\xi}(x_0) \, \delta[x_1 - q(x_0)] \,.
\end{equation}

Performing the (Grassmann) Gaussian integral over the auxiliary fields, we
see that
\begin{equation}
\Gamma[q] \;=\; - {\rm Tr} \ln\big[{\mathcal K}_f\big]\;
\end{equation}
where:
\begin{equation}
{\mathcal K}_f(\tau,\tau') \,=\, \frac{1}{\lambda} \delta(\tau-\tau')
\,+\, {\mathcal S}_f (\tau -\tau', q(\tau)- q(\tau') ) \;.
\end{equation}
We again expand in powers of the fluctuating $q(\tau)$,
\begin{equation}
{\mathcal K}_f \;=\; {\mathcal K}_f^{(0)}  \,+\, {\mathcal K}_f^{(1)} \,+\,
{\mathcal K}_f^{(2)} \,+\,\ldots
\end{equation}
with
\begin{equation}
{\mathcal K}_f^{(0)}(\tau-\tau') \,=\, \int \frac{d\omega}{2\pi} e^{i
\omega (\tau-\tau')} {\widetilde{\mathcal K}}_f^{(0)}(\omega)  \;,\;\;\;
{\widetilde{\mathcal K}}_f^{(0)}(\omega) \,=\, \Big[\frac{1}{\lambda} + \frac{-i
\gamma_0 \omega + m}{2 \sqrt{\omega^2 + m^2}} \Big]\;,
\end{equation}
\begin{equation}
{\mathcal K}_f^{(1)}(\tau-\tau') \,=\, - \gamma_1 \, [q(\tau) - q(\tau')] \,
\int \frac{d\omega}{2\pi} e^{i \omega (\tau-\tau')} \frac{1}{2} \sqrt{\omega^2 + m^2}
\end{equation}
and
\begin{equation}
{\mathcal K}_f^{(2)}(\tau-\tau') \,=\, \frac{1}{4} \, [q(\tau) -
q(\tau')]^2 \, \int \frac{d\omega}{2\pi} e^{i \omega (\tau-\tau')}
\sqrt{\omega^2 + m^2}  \, ( - i \gamma_0 \omega + m) \;.
\end{equation}
Up to second order in the fluctuation,
\begin{eqnarray}
\Gamma[q(\tau)] &\simeq& \Gamma_2[q(\tau)]  \nonumber\\
\Gamma_2[q] &=& -{\rm Tr}\Big[ ({\mathcal K}_f^{(0)})^{-1}{\mathcal K}_f^{(2)}\Big]
+ \frac{1}{2} {\rm Tr}\Big\{ \big[ ({\mathcal
K}_f^{(0)})^{-1}{\mathcal K}_f^{(1)}\Big]^2 \Big\} \nonumber\\
&\equiv& \Gamma_2^{(1)}[q(\tau)]\,+\,\Gamma_2^{(2)}[q(\tau)] \,
\end{eqnarray}
where
\begin{eqnarray}
\Gamma_2^{(1)}[q(\tau)]&=&
-{\rm Tr}\Big[ ({\mathcal K}_f^{(0)})^{-1}{\mathcal
K}_f^{(2)}\Big]\nonumber\\
&=& - \int d\tau \int d\tau' [q(\tau) -
q(\tau')]^2  \, F^{(1)}(\tau-\tau')
\end{eqnarray}
with
\begin{eqnarray}
{\tilde F}^{(1)}(\omega)&=&\int \frac{d\nu}{2\pi}\,
\frac{1}{(\lambda ^2+4) m^2+4 \lambda  \sqrt{m^2+(\omega + \nu)^2} m+(\lambda^2+4)
(\omega + \nu)^2} \nonumber\\
&\times& \Big[ \lambda \sqrt{m^2+\nu^2} (2 m^3+\lambda
\sqrt{m^2+(\omega + \nu)^2} m^2 \nonumber\\
&+& 2 (\omega + \nu)^2 m+\lambda  \nu (\omega +
\nu) \sqrt{m^2+(\omega + \nu)^2})\Big]
\end{eqnarray}
which for the $\lambda \to \infty$ and $m \to 0$ case reduces to:
\begin{equation}
{\tilde F}^{(1)}(\omega)\,=\,\int \frac{d\nu}{2\pi}\,
\frac{(\omega + \nu)}{|\omega + \nu|} \nu |\nu| \;.
\end{equation}
For the remaining term, ${\tilde F}^{(2)}$, a somewhat lengthy calculation
shows that it vanishes (for any value of $\lambda$ and $m$).

For the special case of $m=0$ and and $\lambda \to \infty$, the subtracted
version of ${\tilde F}^{(1)}(\omega)$ is
\begin{equation}
{\tilde F}^{(1)}_s(\omega) \,=\, \frac{2}{3} |\omega|^3 \;.
\end{equation}

\section{Plane mirror coupled to a real scalar field in $d+1$ dimensions}\label{sec:dpo}
We shall consider here the generalization of the calculations of
the previous section, in particular the ones for the real scalar
field, to the case of a flat moving mirror in $d+1$ dimensions.
The mirror's Euclidean world-volume is defined by the equation
\begin{equation}
x_d - q(x_0) \;=\; 0 \;;
\end{equation}
the coordinates $x_1,x_2,\ldots,x_{d-1}$ shall be denoted
collectively by $x_\parallel$, since they are parallel to the
mirror.

In a quite straightforward generalization of the derivation
implemented for the $1+1$ dimensional case, we introduce auxiliary
fields $\xi(\tau,x_\parallel)$, living in $d-1$ dimensions,
obtaining for the $d+1$ dimensional vacuum amplitude, ${\mathcal
Z}^{(d+1)}[q(\tau)]$, the expression:
\begin{equation}
{\mathcal Z}^{(d+1)}[q(\tau)] \,=\, \int {\mathcal D}\xi \; e^{-\frac{1}{2} \int
d\tau \int d\tau' \xi(\tau,x_\parallel) {\mathcal
K}(\tau,x_\parallel;\tau',x_\parallel') \xi(\tau',x_\parallel')} \;,
\end{equation}
where
\begin{equation}
{\mathcal K}(\tau,x_\parallel;\tau',x_\parallel')\,=\,
\frac{1}{\lambda} \delta(\tau-\tau') \delta(x_\parallel - x_\parallel') +
\Delta\big[\tau-\tau',x_\parallel - x_\parallel',q(\tau)-q(\tau')\big] \;.
\end{equation}
The $\xi$-integral is again Gaussian, and we take advantage of the
translation invariance along $x_\parallel$ to Fourier transform with
respect to those coordinates, obtaining
\begin{equation}
\Gamma^{(d+1)}[q(\tau)] \;=\; L^{d-1} \, \int
\frac{d^{d-1}k_\parallel}{(2\pi)^{d-1}} \,
\Gamma^{(1+1)}[q(\tau),m(p_\parallel)]
\end{equation}
where in the last expression we introduced $L^{d-1}$, the `area' of the plate,
and $\Gamma^{(1+1)}[q(\tau),m(k_\parallel)]$ denotes the effective action for
the $1+1$ dimensional case, calculated with a mass depending on the
parallel momentum, through the equation:
\begin{equation}
m^2(k_\parallel) \;=\; m^2 \,+\, k_\parallel^2 \;,
\end{equation}
with $m$ denoting the standard mass of the field.

The $L^{d+1}$ factor is divergent for an infinite plate. This
divergence is, however, harmless from the physical point of view,
since the natural object to calculate is not the force but rather
the {\em pressure\/} experienced by the mirror, hence the area
factor is divided out.

In the quadratic approximation we have
\begin{equation}
\frac{1}{L^{d-1}} \, \Gamma_2^{(d+1)}[q(\tau)] \;=\;
 - \int_{-\infty}^{+\infty} d\tau \int_{-\infty}^{+\infty}  d\tau'
q(\tau) q(\tau') \, F^{(d+1)}_s(\tau -\tau') \;,
\end{equation}
where
\begin{eqnarray}
{\tilde F}^{(d+1)}_s(\omega,\Xi) &=&
\frac{1}{4} \, \int \frac{d^{d-1}k_\parallel}{(2\pi)^{d-1}} \,
\int_{-\Xi}^\Xi \frac{d\nu}{2\pi} \,\Big\{
\big[\frac{1}{\lambda} \,+\, \frac{1}{2 \sqrt{(\nu + \omega)^2 + m^2 +
k_\parallel^2}}\big]^{-1} \nonumber\\
&-& \big[\frac{1}{\lambda} \,+\, \frac{1}{2 \sqrt{\nu^2 + m^2 +
k_\parallel^2}}\big]^{-1} \Big\} \, \sqrt{\nu^2 + m^2 + k_\parallel^2} \;.
\end{eqnarray}
As an example, we consider the particular case $m=0$ and $\lambda
\to \infty$:
\begin{eqnarray}
{\tilde F}^{(d+1)}_s(\omega,\Xi) &=&
\frac{1}{2} \, \int \frac{d^{d-1}k_\parallel}{(2\pi)^{d-1}} \,
\int_{-\Xi}^\Xi \frac{d\nu}{2\pi} \,\Big\{
\sqrt{(\nu + \omega)^2 + k_\parallel^2} \nonumber\\
&-& \sqrt{\nu^2 + k_\parallel^2}\Big\} \,
\sqrt{\nu^2 + k_\parallel^2} \;.
\end{eqnarray}
As already mentioned, the calculation in $d+1$ dimensions is
similar to the massive case in $1+1$ dimensions. Following the
steps described in Appendix A we find
\begin{equation}
\tilde F_s^{(d+1)}(\omega)=
\frac{\Gamma^2(\frac{1+d}{2})\Gamma(-1-(d/2))}{2^{d+3}\pi^{d/2+1}
\Gamma(d+1)}(\omega^2)^{1+\frac{d}{2}}
\label{Fperfd}
\end{equation}

While in an odd number of space dimensions the result could be
predicted by dimensional analysis, there is a subtle point in even
dimensions. As  $\Gamma(-1-(d/2))$ is divergent in this case, it
is necessary to introduce in the Lagrangian a counterterm with
higher derivatives of the mirror's position. Once  the divergence
is absorbed, a finite term remains, proportional to ${\rm
log}[\omega/\mu]$, where $\mu$ is an arbitrary constant,
determined by the renormalization point.

\section{Imaginary part of the in-out effective
action}\label{sec:imaginary}
 One of the most distinctive signals of the dispersive effects
 due to an accelerated mirror is a non-vanishing probability of
 producing a particle pair out of the vacuum.
  Indeed, the total probability of producing a particle
 pair when the whole history of the mirror, from $t \to
-\infty$ to $t \to +\infty$, is taken into account, can be
obtained from the imaginary part of the `in-out' real time
effective action $\Gamma^{io}$:
\begin{equation}
P \;=\; 2 \, {\rm Im}[\Gamma^{io}] \;.
\end{equation}
This real-time effective action can, on the other hand, be
obtained by performing the inverse Wick rotation on the Euclidean
$\Gamma[q(\tau)]$ effective actions that we have just calculated,
back to real time.

Let us obtain the explicit form of $P$ for two illustrative
examples, the cases of the massless and massive real scalar fields
in $1+1$ dimensions, since they encode the main features of the
physical process we want to describe (other cases will indeed give
different results, but they will be kinematical in nature).

The quadratic approximation to the imaginary-time effective
action, $\Gamma_2[q(\tau)]$, whose general form in Fourier space
is:
\begin{equation}
 \Gamma_2[q] \;=\; - \int_{-\infty}^{+\infty} \frac{d\omega}{2\pi} \;
{\tilde F}_s(\omega) \, |{\tilde q}(\omega)|^2 \;,
\end{equation}
leads to
\begin{equation}
 \Gamma_2^{io}[q] \;=\; \int_{-\infty}^{+\infty} \frac{d\omega}{2\pi} \;
{\tilde F}_s(i \omega) \, |{\tilde q}(\omega)|^2 \;,
\end{equation}
where we kept the same notation for the rotated function $q$. Thus,
\begin{equation}\label{eq:P}
 P \;=\; 2 \, \int_{-\infty}^{+\infty}
\frac{d\omega}{2\pi} \; {\rm Im}[ {\tilde F}_s(i \omega) ] \,
|{\tilde q}(\omega)|^2
\;.
\end{equation}
Let us now evaluate $P$ for the two cases mentioned above: in the
simplest case of a massless real scalar field with perfect
boundary conditions in $1+1$ dimensions, we have,
\begin{equation}
{\rm Im} {\tilde F}_s(i\omega) =  \frac{1}{12\pi}
(-\omega^2)^{3/2}=\pm \, \frac{1}{12\pi}  |\omega|^3\;.
\end{equation}
The two signs correspond to the two possible determinations of the
square root. Of course, only the positive one corresponds to the
right physical situation (Feynman conditions):
\begin{equation}\label{eq:P1}
 P \;=\; \frac{1}{6\pi} \, \int_{-\infty}^{+\infty}
\frac{d\omega}{2\pi} \; |\omega|^3 \, |{\tilde q}(\omega)|^2 \;.
\end{equation}

In the case $m \neq 0$ and $\lambda \to \infty$, the imaginary
part of $\tilde F_s$ can also be computed. Indeed, from
(\ref{Fmperf}), we see that
\begin{equation}
{\rm Im} {\tilde F}_s(i\omega) =  \frac{\theta(\omega^2-4
m^2)}{12\pi} \vert \omega^3\vert
\int_0^{1-\frac{4m^2}{\omega^2}}dx\,
\big[1-\frac{4m^2}{\omega^2(1-x^2)}\big]^{3/2}.
\end{equation}
The integral can be computed explicitly in terms of elliptic
functions, but we will not need that rather cumbersome expression
in what follows, since we want to pinpoint a rather interesting
physical phenomenon: the presence of a threshold in the imaginary
part of the effective action. This can be understood as follows:
if the mirror oscillates with a frequency $\omega$, the reflection
of a single field mode, with frequency $\omega_k$, will generate
frequency sidebands $\omega_k-\omega ,\, \omega_k+\omega$. In
order to create particles it is necessary to have a mix of
positive and negative frequencies, and the negative frequency
should be smaller than $-m$, i.e. $\omega_k-\omega < -m$. This
yields $\omega > 2 m$, as predicted by the previous equation.

\section{Real time dynamics: the Schwinger-Keldysh effective
action}\label{sec:eqs}

Up to now, we have considered the Euclidean effective action that
describes the dynamics of the mirror after integration of the
quantum fields, and applied it to study the probability of emitting a
particle pair during the whole evolution of the system, by performing
a Wick rotation back to Minkowski space. The last object is the
in-out effective action, which cannot be applied in a straightforward way
to the derivation of the equations of motion, since they would become
neither real nor causal.

As is well known, in order to get the correct effective equations
of motion, one should compute the in-in, Schwinger-Keldysh or
Closed Time Path Effective Action (CTPEA) \cite{ctp0}, which also has
information on the stochastic dynamics of the mirror \cite{ctp}.
The CTPEA is defined as
\begin{equation} e^{-i \Gamma_{\rm CTP} [q^+,q^-]} =
\int {\cal D} \phi^+{\cal D} \phi^- e^{i(S[q^+,\phi^+]-
S[q^-,\phi^-])}, \label{ctpeff}\end{equation} and the field
equations are obtained taking the variation of this action with
respect to the $q^+$, and then setting $q^+=q^-$. As in the
Euclidean case Eq.(\ref{auxfield}), one can introduce two
auxiliary fields $\xi_\pm(t)$ living in $0+1$ dimensions, in order
to linearize the coupling between the mirror and the field.
Instead of doing this, we will follow an alternative procedure.
Using a more concise notation, we can write the CTPEA as
\begin{equation} e^{- i \Gamma_{\rm CTP}[q]} =
\int {\cal D} \phi e^{i S^{{\cal
C}}[q,\phi]},\label{neweff}\end{equation} where we have introduced the CTP
complex temporal path ${\cal C}$, going from minus to plus infinity $\cal C_+$
and backwards $\cal C_-$, with a decreasing (infinitesimal) imaginary part.
Time integration over the contour ${\cal C}$ is defined by $\int_{{\cal C}} dt
=\int _{{\cal C_+}} dt -\int_{{\cal C_-}} dt$. The field $\phi$  appearing in
Eq.(\ref{neweff}) is related to those in Eq.(\ref{ctpeff}) by $\phi(t,\vec x) =
\phi_{\pm}(t,\vec x)$ if $t \in {\cal C}_{\pm}$. The same applies to
the mirror's position $q$.

The equation above is useful because it has the structure of the
usual in-out or the Euclidean effective action. Feynman rules are
therefore the ordinary ones, replacing Euclidean propagator by
\cite{ctprules}
\begin{eqnarray} G(x,y) = \left\{\begin{array}{ll}
G_F(x,y)=i \langle 0, in\vert T \phi (x) \phi(y)\vert 0, in\rangle,& ~t, t' ~
\mbox{both on} ~{\cal C}_+ \\ G_D(x,y)=-i \langle 0, in\vert {\tilde T}\phi (x)
\phi(y)\vert 0, in\rangle , & ~t, t' ~ \mbox{both on} ~{\cal C}_-\\ G_+(x,y)=-
i \langle 0, in\vert \phi (x) \phi(y)\vert 0, in\rangle,    &~t ~\mbox{on}~
{\cal C}_-, t'  ~\mbox{on} ~{\cal C}_+\\ G_-(x,y)=i \langle 0, in\vert \phi (y)
\phi(x)\vert 0, in\rangle, & ~ t  ~\mbox{on} ~ {\cal C}_+, t'~  \mbox{on}~
{\cal C}_-\end{array}\right. \label{prop}
\end{eqnarray}
Explicitly
\begin{equation} G_F(x,y)= \int \frac{d^{d+1}p}{(2 \pi)^{d+1}}\frac {e^{ip(x-
y)}}{p^2+m^2-i\epsilon}= G_D^*(x,y),
\end{equation}
\begin{equation}
G_{\pm}(x,y)=\mp \int \frac{d^{d+1}p}{(2 \pi)^{d+1}} e^{ip(x-y)}2
\pi i
\delta(p^2-m^2)\theta(\pm p^0).
\label{gpm}
\end{equation}

The considerations above will allow us to compute the CTPEA using
the Euclidean results of the previous sections. To do this, we
will rewrite the Euclidean effective action given in Eq.(\ref{eea})
using a spectral decomposition for the form factor $F_s(\tau -
\tau ')$. For definiteness we will consider the concrete example
of a dispersive mirror (finite $\lambda$) coupled to a massless
scalar field.

The Euclidean effective action can be
rewritten as
$$\Gamma_2[q(\tau )]= -\frac{\lambda^2}{4\pi^3}\int d\tau\int d\tau'q(\tau )
q(\tau')\int_0^{\infty}dz \left(1 -
f(2z/\lambda)\right)\frac{d^2}{d\tau^2}G_{\rm E}(\tau -
\tau',z^2),$$ where $G_E(\tau,z^2)$ is 0+1 Euclidean propagator
with mass $z^2$ and
\begin{equation}
f(z) = \frac{\arctan z}{z}+ \frac{1}{2}\ln (1 + z^2)
\end{equation}
(see Appendix B for details).

The CTPEA can be obtained from the Euclidean one
considering the contour ${\cal C}$ and replacing $G_E(z,t)$
according to the rules given in Eq.(\ref{prop}). The result is
\begin{eqnarray}
 \Gamma_{\rm CTP} &=& \frac{\lambda^2}{4\pi^3}\int dt\int dt' q^{\rm a}(t)
 q^{\rm b}(t')\int_0^{\infty}dz(1 - f(2z/\lambda)) \psi_{\rm
 ab}\nonumber \\
&=&\frac{\lambda^2}{4\pi^3}\int_0^{\infty}dz(1 -
f(2z/\lambda))\left[\int dt\int dt'q^+(t) q^+(t')
\psi_{++}
-\int dt\int dt'q^+(t) q^-(t') \psi_{+-}\right.\nonumber \\
&+&\int dt\int dt'q^-(t) q^+(t') \psi_{-+}
-\left.\int dt\int dt'q^-(t) q^-(t') \psi_{--}\right],
\end{eqnarray} where the CTP propagators are
\begin{eqnarray} \psi_{\pm\pm}(t,t')&=& \pm
\frac{d^2}{dt^2}\int\frac{d\omega}{2\pi}\frac{e^{i\omega(t -
t')}}{\omega^2 - z^2 \pm i\epsilon},\nonumber\\ \psi_{\pm\mp}(t,t')&=&
\mp \frac{d^2}{dt^2}\int\frac{d\omega}{2\pi}e^{i\omega(t - t')}
2\pi i \delta (\omega^2 - z^2)\theta (\pm \omega),
\end{eqnarray}
and satisfy the identities
\begin{eqnarray}
 \psi_{++} &=&
\psi_{+-} \theta (t - t') - \psi_{-+}\theta
(t'-t),\nonumber\\
\psi_{--} &=& \psi_{-+} \theta (t - t') - \psi_{+-}\theta (t'-t).
\end{eqnarray}

Introducing the new variables $\Sigma = (q^++q^-)/2$ and $\Delta =
(q^+-q^-)/2 $, the CTP effective action can be written as
\begin{equation}\Gamma_{\rm CTP} = \int dt \int dt' \left[ \ddot{\Sigma}(t) \Sigma(t')
D(t'-t) - i \ddot{\Delta}(t) \Delta (t') N(t-t')\right],
\label{ctpeafin}
\end{equation}
where the dissipation $(D)$ and noise $(N)$ kernels are given by
\begin{eqnarray}D(t-t')&=& \frac{2\lambda^2}{\pi^2}\int_0^{\infty}dz~\left(1 -
f(2z/\lambda)\right) ~{\rm Re}\psi_{++}(t,t') ~\theta (t'-t),\nonumber\\
N(t-t')&=& \frac{\lambda^2}{\pi^2}\int_0^{\infty}dz~ \left(1 -
f(2z/\lambda)\right) ~{\rm Im}\psi_{++}(t,t'). \label{kernels}
\end{eqnarray}
Performing the integral in the spectral parameter $z$ we find
\begin{equation}
D(r) = \frac{\lambda^2}{2\pi}\left[(\frac{r \lambda}{2}+1) {\rm
ChI}(\frac{r\lambda}{2}) + \cos(\frac{r\lambda}{2})-
\sinh(\frac{r\lambda}{2}) -(\frac{r \lambda}{2}+1) {\rm
ShI}(\frac{r\lambda}{2})\right],
\end{equation}
and \begin{eqnarray} N(r) &=& \frac{-\lambda^2}
{48\pi^2}\left[-\pi^2 -24\gamma \ln (\frac{2}{\lambda}) + 12
\ln^2(2) - 12 \ln^2(\lambda) - 24
\ln(\frac{2}{\lambda})\left(\ln(r\lambda) - {\rm
ChI}(\frac{r\lambda}{2})\right)\right. \nonumber \\
&+& \left. 24 (\ln(\frac{2}{\lambda})-1) \left(\cosh
(\frac{r\lambda}{2}) - \frac{r\lambda}{2}{\rm
ShI}(\frac{r\lambda}{2})\right) \right]\end{eqnarray} where $r =
\vert t - t'\vert $, and ${\rm ChI}$, ${\rm ShI}$ are the
hyperbolic CosIntegral and SinIntegral, respectively.

From Eq.(\ref{ctpeafin}) we can also see that the dissipative force is given
by
\begin{equation}
F_{\rm diss}=\int_{-\infty}^t dt' \ddot q(t') D(t-t').
\end{equation}
Taking into account that dissipation kernel has the form $D(t-t')=
\lambda^2 g(\lambda (t-t'))$, we can rewrite the force as
\begin{equation}
F_{\rm diss}=\lambda \int_{0}^\infty dx \, D(x) \ddot
q(t-\frac{x}{\lambda}) \simeq \lambda \int_{0}^\infty dx \, D(x)
\left [\ddot q(t) -\frac{x}{\lambda}q^{(3)}(t)+....\right].
\end{equation}
A numerical evaluation of the remaining integrals gives the
correct perfect conductor limit: the term proportional to
$\lambda$ renormalizes the mass of the mirror, and the
$\lambda$-independent term gives a dissipation force proportional
to the third derivative of the mirror's position.

In order to derive the semiclassical Langevin equation that describes the
motion of the mirror, one can regard the imaginary part of $\Gamma_{\rm CTP}$
as coming from a noise source $\eta(t)$ with Gaussian functional
probability distribution given by

\begin{equation}
P[\eta (t)] = N_\eta \exp\left\{-\frac{1}{2}\int dt\int dt' \left[\eta (t) {\ddot N}^{-1}(t - t')
\eta (t')\right]\right\},
\end{equation}
where $N_\eta$ is a normalization factor. Indeed, we can write the
imaginary part of the CTP-effective action as a functional
integral over the Gaussian field $\eta(t)$

\begin{equation}
\int {\cal D}\eta (t) P[\eta ] ~e^{-i \Delta (t) \eta (t)} = e^{-i \int dt \int dt' \Delta (t)
{\ddot N}(t - t') \Delta (t')}.
\end{equation}
Therefore, the CTPEA can be rewritten as an average over the noise
of

 \begin{equation}\Gamma_{\rm CTP}^{(\eta)} = \int dt \int dt' \left[ \ddot{\Sigma}(t) \Sigma(t')
D(t'-t)\right] - \int dt  \Delta (t) \eta (t). \label{ctpeafin2}
\end{equation}

Thus, the associated Langevin equation comes from the variation
$\frac{\delta \Gamma_{\rm CPT}^{(\eta)}} {\delta
q_+}\vert_{q_+=q_-}=0$, obtaining

\begin{equation}
M\left( {\ddot q}(t) + \Omega^2 q^2(t)\right) + 2 \int dt' {\ddot
D}(t - t') q(t') = \eta (t), \label{langevineq}
\end{equation}
where $M$ is the mass of the mirror and the two-point correlation
function of the noise is given by

\begin{equation}
\langle \eta (t) \eta (t')\rangle = {\ddot N}(t - t').
\end{equation}
The Langevin equation describes the motion of the system (the
mirror) taking into account the main effects of the environment (the
quantum field): a dissipative force and a stochastic noise.

\subsection{Mirror's decoherence}

In the quantum open system approach that we have adopted here,
the imaginary part of the CTP-effective action (noise term) is
directly associated with the decoherence process of the mirror.
In fact, one can establish a direct link between the total number
of created particles and the decoherence functional
for a given classical (macroscopic) trajectory of the
mirror.

Decoherence means physically that the different coarse-graining
histories making up the full quantum evolution acquire individual
reality, and may therefore be assigned definite probabilities in
the classical sense. For our particular application, we wish to
consider as a single coarse-grained history all those fine-grained
ones where the trajectory $q(t)$ remains close to a prescribed
classical configuration $q_{\rm cl}$.

In principle, we can examine adjacent general classical solutions
for their consistency but, in practice, it is simplest to restrict
ourselves to particular solutions $q^{\pm}_{\rm cl}$, according to
the nature of the decoherence that we are studying. Therefore, we
evaluate the decoherence functional \cite{DF} for classical
trajectories such that the amplitude of one trajectory is $q^- =
q^+ - 2 \delta$, where $\delta$ is a small (constant) amplitude
difference. Thus, neglecting the dissipation we can write
$\Delta_{\rm cl}(t) = \delta \cos(\Omega t)$, and the decoherence
functional is formally given by

\begin{equation}
\vert{\cal D}(q_{\rm cl}^+,q_{\rm cl}^-)\vert =e^{-{\rm
Im}\Gamma_{\rm CTP}}=e^{-\delta^2 \Omega^2 \int dt \int dt' ~
\cos(\Omega t)N(t -t') \cos(\Omega t')}.
\end{equation}
Making the integration in the particular case $m=0$ and
$\lambda=\infty$, one can show that the decoherence time scales as

\begin{equation}
t_D \sim \frac{1}{\delta^2\omega^3}.
\end{equation}

This result is valid as long as the decoherence time is much shorter
than the dissipative time $t_{diss}$, which can be easily estimated
from Eq.(\ref{langevineq})  as $t_{\rm diss}\sim M/\Omega^2$. The
condition $t_D\ll t_{\rm diss}$ is satisfied as long as $\delta\gg
\sqrt{M\Omega}$, i.e the minimum uncertainty in the position of the
mirror. An alternative estimation based on the Fokker-Planck
equation for the Wigner function of the mirror gives the same order
of magnitude for the decoherence time $t_D$ \cite{paulodiego}.

\section{Final remarks}
 In this paper we have extended, in several directions, the functional
approach to the dynamical Casimir effect introduced some years ago by
Golestanian and Kardar~\cite{GK} to consider different situations.
The main point in this approach, namely, the introduction of auxiliary
fields in the functional integral to impose the boundary conditions on the
quantum fields is retained, altough now they have an extra piece in the
action, to cope with the dispersive nature of the mirror.

After integration of the original quantum fields, the problem is
again reduced to the computation  of a path integral over the
auxiliary fields. This is a kind of (non local) dimensionally reduced
theory, since the auxiliary field live on the boundary.

Firstly, we considered non-perfectly conducting
mirrors, by introducing a $\delta$-like potentials for the quantum
fields. As shown in Ref.\cite{barton1} for the scalar field, these
potentials serve as toy models to describe the interaction of the
electromagnetic field with a thin plasma sheet, and give rise to reflection
and transmission coefficients with a particular frequency dependence.
We believe that this generalization will be useful as a first step
towards solving more realistic situations. Indeed, our formalism can be
extended to include arbitrary reflection and transmission coefficients by
considering non local extensions of the singular potentials
considered here. We will describe this results in a forthcoming
publication.

We also considered a scalar field coupled to a relativistic mirror,
and also calculated the effective action for a mirror interacting with a
Dirac spinor, understanding here by mirror an object that reflects the
fermionic current.

Finally, we also extended the formulation to calculate the CTP effective
action. As an important by-product, we applied this  effective action to
compute the semiclassical Langevin equation that describes the dynamics of
the mirror interacting with the vacuum fluctuations of the quantum fields,
and with the motion induced radiation produced by its accelerated motion.

\section*{Appendix A: Massive case}
We outline here the calculation of the Euclidean effective action
in the massive case, for perfectly conducting mirrors in $1+1$
dimensions.

In the definition of ${\tilde F}(\omega)$,
\begin{equation}
\tilde F(\omega)=\frac{1}{2}\int_ 0^\infty
\frac{d\nu}{2\pi}\sqrt{\nu^2+m^2}\sqrt{(\nu+\omega)^2+m^2} \;,
\end{equation}
we insert the representation
\begin{equation}
(\nu^2+m^2)^{\epsilon}=\frac{1}{\Gamma(-\epsilon)}\int_0^{\infty}\frac{d\beta}{\beta}
\beta^{-\epsilon}e^{-\beta(\nu^2+m^2)},
\end{equation}
and one with a shifted argument for the second factor, to obtain
the function
\begin{equation}
\tilde F_\epsilon(\omega)=
\frac{1}{2[\Gamma(-\epsilon)]^2}\int_0^\infty\frac{d\alpha_1}{\alpha_1}
\int_0^\infty\frac{d\alpha_2}{\alpha_2}\alpha_1^{-\epsilon}\alpha_2^{-\epsilon}
\int_ 0^\infty
\frac{d\nu}{2\pi}\exp\big[-\alpha_1(\nu^2+m^2)-\alpha_2((\nu+w)^2+m^2)\big]\;.
\end{equation}
The role of this representation is, to make it possible to
integrate over the frequency; in a way, it allows for the
introduction of Feynman-like parameters in the case of propagators
which have a non-standard form. Indeed, introducing the identity
\begin{equation}
1=\int_0^\infty d\rho\, \delta(\rho-\alpha_1-\alpha_2)\,,
\end{equation}
and after a rescaling of the $\alpha$'s plus some straightforward
calculations we see that
\begin{equation}
\tilde F_\epsilon(\omega)=\frac{\Gamma(-2\epsilon-1/2)}{4\sqrt{\pi}[\Gamma(-\epsilon)]^2}
\int_0^1d\alpha [\alpha(1-\alpha)]^{-1-\epsilon}[\alpha(1-\alpha)\omega^2+m^2]^{2\epsilon+1/2}
\end{equation}
The final result (\ref{Fmperf}) is obtained by subtracting $\tilde
F_\epsilon(0)$ and then taking the limit $\epsilon =1/2$. Note
that the previously used representation is not used as a
regularization, but as a device to do the integral. Indeed, one
could have worked with $\epsilon=1/2$ throughout in the subtracted
integral; no analytic extension to complex values of $\epsilon$
would be required.

The result for the massless case in $d+1$ dimensions,
(\ref{Fperfd}), can be derived using a similar procedure, but now
introducing a dimensional regularization for the momentum
integral. Now a regularization {\em is\/} required, since the
integral over the parallel momenta is divergent when $\epsilon =
1/2$.

\section*{Appendix B: Spectral decomposition}

In this Appendix we derive the spectral decomposition which allows us
to write the form factor $F(\tau - \tau')$ in terms of the Euclidean
propagator. Using the identities
\begin{equation}\vert\omega\vert  =
\frac{2\omega^2}{\pi}\int_0^{+\infty}dz \frac{1}{\omega^2 +
z^2},\end{equation}
\begin{equation}\left(1 + \frac{2}{\lambda}\vert\omega\vert\right)\ln\left(1 +
\frac{2}{\lambda}\vert\omega\vert\right)  =
\omega^2\frac{8}{\lambda^2\pi}\int_0^{+\infty}\frac{f(z)}{z^2 +
\frac{4\omega^2}{\lambda^2}}\,\, , \label{ident}
\end{equation}
where
\begin{equation}
f(z) = \frac{\arctan z}{z}+ \frac{1}{2}\ln (1 + z^2)\,\, ,
\end{equation}
we can write
\begin{equation}
F(\tau - \tau') = -\frac{d^2}{d\tau^2} G(\tau - \tau')
\end{equation}
with \begin{eqnarray} G(\tau) &=& \int
\frac{d\omega}{2\pi}e^{i\omega \tau} {\tilde G}(\omega
)\nonumber\\
{\tilde G}(\omega ) &=&
\frac{\lambda^2}{4\pi^3}\int_0^{+\infty}dz\left\{\frac{1}{\omega^2
+ z^2}\left[1 - f\left(\frac{2z}{\lambda}\right)\right]\right\}.
\end{eqnarray}
From these equations the form factor $G(\tau )$ can readily be
written in terms of the $0+1$ Euclidean propagator $G_{\rm
E}(\tau,z^2)$ with mass $z^2$
\begin{equation}
G(\tau )=\frac{\lambda^2}{4\pi^3}\int_0^{+\infty}dz\left[1 -
f\left(\frac{2z}{\lambda}\right)\right] G_{\rm E}(\tau, z^2)
\end{equation} where
\begin{equation}G_{\rm E}(\tau, z^2) = \int \frac{d\omega}{2\pi}
\frac{e^{i\omega\tau}}{\omega^2+z^2}.
\end{equation}

\acknowledgments
C.D.F. thanks CONICET and ANPCyT for financial support and to the FCEyN
(UBA) by the hospitality of its members.
F.D.M. acknowledges the warm hospitality of Centro At\'omico Bariloche,
where part of this work was done. The work of F.D.M. and F.C.L was
supported by UBA, CONICET and ANPCyT.
%%%%%%%%%%%%%%%%%%%%%%%%%%%%%%%%%%%%%%%%%%%%%%%%%%%%%%%%%%%%%%%%%%%%%%%%%
%%%%%%%%%%%%%%%%%%%%%%%%%%% References %%%%%%%%%%%%%%%%%%%%%%%%%%%%%%%%%%
%%%%%%%%%%%%%%%%%%%%%%%%%%%%%%%%%%%%%%%%%%%%%%%%%%%%%%%%%%%%%%%%%%%%%%%%%

\end{document}